\documentclass{optica-article}

\journal{opticajournal} % for journals or Optica Open

\articletype{Research Article}

\usepackage{lineno}
\begin{document}

\title{Quantum teleportation coexisting with classical communications in optical fiber}

\author{Jordan M. Thomas,\authormark{1,*} Fei I. Yeh,\authormark{2}
Jim Hao Chen, \authormark{2} 
Joe J. Mambretti, \authormark{2}
Scott J. Kohlert, \authormark{3}
Gregory S. Kanter,\authormark{1,4} and Prem Kumar\authormark{1,5}}

\address{\authormark{1}Center for Photonic Communication and Computing, Northwestern University, 2145 Sheridan Road, Evanston, IL 60208, USA\\
\authormark{2}International Center for Advanced Internet Research, Northwestern University, 750 N. Lake Shore Drive, Chicago, IL 60611, USA\\
\authormark{3}Ciena Corporation, 7035 Ridge Road, Hanover, MD 21076, USA\\
\authormark{4}NuCrypt, LLC, 1460 Renaissance Drive, Suite \#205, Park Ridge, IL 60068, USA\\
\authormark{5}Department of Physics and Astronomy, Northwestern University, 2145 Sheridan Road,
Evanston, IL 60208-3112, USA}

\email{\authormark{*}jordanthomas2025@u.northwestern.edu} %% email address is required; see note below about the corresponding author designation

% use {asbstract*} to suppress the copyright line. Copyright information will be added in production

\begin{abstract*} 
The ability for quantum and conventional networks to operate in the same optical fibers would aid the deployment of quantum network technology on a large scale. Quantum teleportation is a fundamental operation in quantum networking, but has yet to be demonstrated in fibers populated with high-power conventional optical signals. Here we report to the best of our knowledge the first demonstration of quantum teleportation over fibers carrying conventional telecommunications traffic. Quantum state transfer is achieved over a 30.2-km fiber carrying 400-Gbps C-band classical traffic with a Bell state measurement performed at the fiber’s midpoint. To protect quantum fidelity from spontaneous Raman scattering noise, we use optimal O-band quantum channels, narrow spectro-temporal filtering, and multi-photon coincidence detection. Fidelity is shown to be well maintained with an elevated C-band launch power of 18.7\,dBm for the single-channel 400-Gbps signal, which we project could support multiple classical channels totaling many terabits/s aggregate data rates. These results show the feasibility of advanced quantum and classical network applications operating within a unified fiber infrastructure.
\end{abstract*}

%%%%%%%%%%%%%%%%%%%%%%%%%%  body  %%%%%%%%%%%%%%%%%%%%%%%%%%
\section{Introduction}

The optical fiber infrastructure and telecommunications technology that underlie the Internet have been widely adopted by researchers seeking to develop quantum networks capable of applications such as quantum-enhanced cryptography, sensing, and networked quantum computing \cite{kimble_quantum_2008, wehner_quantum_2018}. However, since the
majority of the existing fiber infrastructure is populated with conventional telecommunications traffic and due to the economic cost of leasing or installing new fiber, whether quantum networking can be realized on a large scale will depend on the ability to propagate quantum signals in the same fiber as high-power classical signals.\\
\indent Quantum and classical signals can readily share a single fiber via wavelength division multiplexing. However, it was first shown in \cite{townsend_simultaneous_1997} that noise photons from inelastic scattering of the high-power classical light can obscure the detection of often sub-photon-level quantum signals. Of the potential sources of noise, spontaneous Raman scattering (SpRS) is most dominant due to its broadband spectrum meaning noise can never be entirely prevented \cite{chapuran_optical_2009}. Without careful design, a trade-off between the capabilities of conventional and quantum networking arises.\\
\indent The study of quantum-classical coexistence has a long history with many experiments conducted for applications using weak coherent state (WCS) sources \cite{townsend_simultaneous_1997, chapuran_optical_2009, eraerds_quantum_2010, qi_feasibility_2010, patel_coexistence_2012, dynes_ultra-high_2016, PhysRevA.95.012301, Mao:18, Geng:21, gavignet_co-propagation_2023}, entangled photon pairs \cite{1637099,ciurana_entanglement_2015,10.1117/12.2230222, kapoor_picosecond_2023, Thomas:23,CCband_ent,hyper,coex_ecoc23, Qwrap,NIST_coex}, Bell state measurements (BSMs) on WCSs \cite{valivarthi_measurement-device-independent_2019, berrevoets_deployed_2022}, continuous variables (CVs) \cite{kumar_coexistence_2015}, and squeezed light \cite{Chapman:23}. Although significant achievements and insights have been made in these studies, all experiments to date have focused on systems that directly transmit quantum information between network nodes. However, many next-generation quantum applications require the disembodied transfer of quantum states between users. Enabled by the non-local properties of quantum entanglement, quantum teleportation allows the transfer of a quantum state between two distant physical systems without the need for direct transmission \cite{PhysRevLett.70.1895}. It plays a foundational role in advanced applications such as quantum relays \cite{relay}, quantum repeaters \cite{PhysRevLett.81.5932}, networking quantum computers \cite{PhysRevLett.96.010503}, and other applications in quantum science and technology \cite{hu_progress_2023}. \\
\indent Recent years have seen impressive progress towards implementing teleportation-based networks in quantum-dedicated fiber \cite{swisscom,valivarthi_quantum_2016, sun_quantum_2016, sun_entanglement_2017-2, valivarthi_teleportation_2020, QFC1, QFC2, shen_hertz-rate_2023, hu_progress_2023}, but teleportation over fibers carrying conventional communications has yet to be demonstrated. As a result, many open questions on the feasibility, limitations, and potential benefits of teleportation in quantum-classical networking remain unexplored.\\
\indent In this paper, we demonstrate a three-node quantum state teleportation system operating over 30.2\,km of optical fiber that simultaneously carries high-power C-band classical communications at a rate of 400\,Gbps. Teleportation is performed via a joint BSM on a single photon and one member of an entangled Bell state photon pair \cite{PhysRevA.51.R1727}. 
Utilizing wavelength multiplexing of quantum and classical signals, a heralded single photon carrying a polarization encoded qubit and one photon of an entangled Bell state photon pair are distributed over roughly 15\,km of spooled fiber to undergo a BSM near the midpoint of the fiber, whilst classical light traverses the full 30.2-km link plus an additional 48\,km of deployed fiber.\\
\indent We implement multiple techniques to mitigate degradation due to noise photons. At the quantum source nodes, we make use of the low SpRS noise in the O-band from strong C-band light \cite{townsend_simultaneous_1997}, which has been shown successful for protecting quantum states from high-power C-band communications \cite{PhysRevA.95.012301, Mao:18, Geng:21,gavignet_co-propagation_2023,berrevoets_deployed_2022,Thomas:23}. More specifically, we allocate coexisting photons to the 1290-nm quantum channels to have nearly the lowest SpRS generation probability achievable using telecom-band wavelength engineering \cite{Thomas:23}. At quantum receivers, we use narrow-band spectro-temporal filtering to reject the detection of uncorrelated SpRS photons. Furthermore, each source uses spontaneous parametric down-conversion to generate photon pairs with strong temporal correlations to further reject noise via coincidence detection of the four photons.\\
\indent We first demonstrate that our approach allows high-power C-band communications to have no significant impact on entanglement distribution to the BSM node nor Hong-Ou-Mandel (HOM) interference \cite{HOM} between independent heralded photons, which are the underlying operations of the quantum teleportation protocol \cite{PhysRevA.51.R1727}. We then perform quantum state teleportation and demonstrate that fidelity is maintained at elevated C-band launch powers as high as 74\,mW (18.7\,dBm), which is on the order of the highest achieved using direct transmission of quantum states  \cite{Mao:18,Geng:21,gavignet_co-propagation_2023,Thomas:23,kumar_coexistence_2015}. Using a theoretical model, we then discuss the the pros and cons of using noisier quantum wavelength channels (e.g., C-band), the effects of storage in imperfect quantum memories, and compare to methods using direct qubit transmission. \\
\indent This experiment demonstrates approaches for integrating many key classical and quantum network components, including high-rate classical signals, optical amplifiers, single and entangled photon pair distribution, and multi-qubit operations such as BSMs all operating simultaneously in the same fibers.

\begin{figure*}[!t]
\centering
\includegraphics[width=\linewidth]{fig1finalfinalfinal.pdf}
\caption{(a) Conceptual diagram of the experiment. Alice prepares a qubit $|\psi\rangle_A$ that she wishes to transfer onto a photon at Bob's node via quantum state teleportation. Alice encodes $|\psi\rangle_A$ onto a single photon, which is wavelength-division multiplexed into a fiber of length $L_{AC}$ to co-propagate with conventional classical communications traffic to a node at Charlie. The classical signal is de-multiplexed to bypass Charlie's node and re-multiplexed to continue propagating over a fiber of length $L_{BC}$. Two photons generated at Bob's node are prepared in an entangled Bell state, where one is transmitted to counter-propagate with the classical light to Charlie to undergo a Bell state measurement (BSM) with Alice's qubit. After a BSM detection, Alice's state is destroyed whilst the photon kept at Bob's node is projected onto the state $\sigma |\psi\rangle_A$, where $\sigma$ is a unitary Pauli transformation identifiable from the BSM result. (b) Experimental implementation. Quantum and classical signals coexist in 30.2\,$\text{ km}$ of spooled fiber ($L_{AC} = 15.2 \text{ km}$, $ L_{BC} = 15.0 \text{ km}$). Alice and Bob both generate photon pairs via non-degenerate spontaneous parametric down-conversion with center wavelengths of 1290\,nm and 1310\,nm. Alice's source heralds a single photon, whereas Bob prepares a two-photon entangled Bell state. Qubits are encoded in polarization and the 1290-nm photons from each source are transmitted to Charlie for the BSM. Teleportation is evaluated via four-fold coincidence detection and polarization analysis of Bob's target photon. The classical source transmits a 400-Gbps C-band signal (1547.32 nm) over 24 km of deployed fiber before being amplified and multiplexed into the 30.2-km spool. After de-multiplexing this propagates over another 24-km deployed fiber to the receiver, totaling 78.2 km of fiber. (FPC = fiber polarization controller, DWDM = dense wavelength division multiplexer, FWDM = O-band/C-band WDM, PPLN = periodically poled lithium niobate waveguide, $\lambda/2$ = half-wave plate, $\lambda/4$ = quarter-wave plate, LP = linear polarizer, LCR = liquid crystal retarder, PBS = polarizing beam splitter, FBS = 50:50 fiber coupler, VDL = variable optical delay line, FBG = fiber Bragg grating, CIR = circulator, $\text{D}_j$ = superconducting nanowire single photon detector, C = common port, P = pass port, R = reflect port, EDFA = erbium doped fiber amplifier).}
\label{fig:fig1}
\end{figure*}

\section{Experimental design}
A conceptual diagram of the experiment is shown in Fig. \ref{fig:fig1}(a). Alice generates a single photon to encode the quantum state $|\psi\rangle_A = \alpha |0\rangle + \beta |1\rangle$ that she wishes to teleport to Bob. Alice’s photon is multiplexed into an optical fiber of length $L_{AC}$ to co-propagate with a classical communications signal to a BSM node at Charlie. The classical signal is de-multiplexed just prior to the BSM and then re-multiplexed to bypass Charlie's node. The classical signal then travels along another fiber of length $L_{BC}$, counter-propagating with respect to one photon from an entangled Bell state photon pair generated at Bob’s node. During the BSM at Charlie, both photons are irreversibly destroyed by the detection whilst Bob’s other photon of the entangled pair is projected onto the state $|\psi\rangle_B = \sigma |\psi\rangle_A$, where $\sigma$ is a unitary operation that is unique to the BSM result and can be classically communicated to Bob to recover Alice's state \cite{PhysRevLett.70.1895} either by physically applying the unitary or accounting for it in data post-processing. Fig. \ref{fig:fig1}(b) shows the physical realization of the experiment. In our case, Alice's state is encoded onto a heralded single photon's polarization, $L_{AC}$ is 15.2 km, and $L_{BC}$ is 15.0 km. Thus, the quantum state transfer distance, and the distance in which quantum and classical signals coexist, is 30.2 km.\\
\indent The classical source consists of a 400-Gbps C-band transceiver operating at 1547.32\,nm (Ciena WaveLogic 5 Nano 400ZR). To reduce the amount of SpRS noise photons that are generated into the quantum channels, we generate photons at O-band wavelengths so that they have an anti-Stokes frequency detuning from the C-band light as well as suppressed Raman gain at the far offset detuning between the C- and O-bands \cite{chapuran_optical_2009}. Instead of the most commonly used 1310-nm O-band channel \cite{PhysRevA.95.012301, Mao:18, Geng:21,gavignet_co-propagation_2023,berrevoets_deployed_2022}, we suppress SpRS by roughly another order of magnitude using the 1290-nm channel due to the multi-mode SpRS noise spectrum from C-band light \cite{Thomas:23}.\\
\indent Both Alice and Bob generate their respective quantum signals via type-0 cascaded second harmonic generation-spontaneous parametric down conversion (c-SHG-SPDC) photon pair generation \cite{Arahira}. Alice and Bob each have independent periodically-poled lithium niobate (PPLN) waveguides that are phase matched for SHG at 650\,nm. The cascaded second-order nonlinear process is analogous to four-wave mixing, which we pump using a 1300-nm continuous-wave distributed feedback laser that is intensity modulated using a lithium niobate Mach-Zehnder modulator to generate pulses with an approximate 65-ps temporal full-width at half maximum (FWHM) and a pulse repetition rate of 500\,MHz. This pump is then split and directed to Alice and Bob's nodes (see supp. materials). The 650-nm pulses generated inside each waveguide simultaneously drives SPDC to produce a superposition of photon pairs with a non-degenerate spectrum centered around 1300 nm and forms the basis for Alice's heralded single-photon source (HSPS) and generating Bob's entangled photons, but requires a low single-photon pair generation rate to minimize the probability of multi-photon emission \cite{Takeoka_2015}\\
\indent Alice's qubit is initially encoded onto a heralded single photon. Her PPLN waveguide is pumped directly to generate photon pairs. After the waveguide, we use a 100-GHz 1300-nm dense wavelength division multiplexer (DWDM) as a notch filter for the pump, followed by a 1310-nm DWDM to separate each photon into different fibers via the DWDM's pass and reflect ports. The 1310-nm photon is filtered by
a fiber Bragg grating (FBG) with FWHM bandwidth of 60\,pm and then detected by a superconducting nanowire single-photon detector (SNSPD). Due to energy conservation, this detection heralds the presence of a photon with a center wavelength of 1290\,nm in the other mode. Alice's qubit is then encoded onto the horizontal (\textit{H}) and vertical (\textit{V}) components of the heralded photon's polarization $|\psi\rangle_A = \alpha |H\rangle + \beta|V\rangle$ using polarization waveplates. This photon is then multiplexed using an O-band/C-band WDM to co-propagate with the 400-Gbps C-band classical signal over 15.2\,km of spooled optical fiber (SMF-28(R) ULL) to Charlie. \\
\indent Bob's source is designed similarly, except his PPLN waveguide is placed inside a polarization Sagnac loop to generate polarization entangled photon pairs using c-SHG-SPDC \cite{Arahira}. Bob's 1290-nm photon is transmitted to counter-propagate with the 400-Gbps signal over 15.0\,km of spooled fiber (SMF-28(R) ULL) to Charlie. However, Bob's 1310-nm photon is kept locally to act as the target photon for teleporting Alice's state to Bob's node. This photon is sent to a free-space polarization module which compensates polarization rotations in the fibers and prepares Bob's photons in the $|\Psi^-\rangle_B$ Bell state using alignment signals and coincidence detection \cite{alignment}, and then uses a quarter-waveplate, half-waveplate, and polarizing beam splitter (PBS) combination to characterize the polarization of Bob's target photon. Subsequently, this photon is filtered by a 56\,pm bandwidth FBG filter and then detected by an SNSPD.\\
\indent At Charlie's node, a BSM is performed on Alice and Bob's incident polarization qubits using conditional HOM interference at a 50:50 beam splitter \cite{PhysRevA.51.R1727}. Since HOM interference only occurs between identical photons, the fidelity of the BSM is governed by their spatial and spectro-temporal indistinguishability. To achieve this, a single-mode fiber 50:50 coupler is used for spatial indistinguishability and a variable optical delay in Alice's pump path is used to ensure that both photons arrive simultaneously at the coupler. Furthermore, the 1290-nm photons are filtered by 60-pm bandwidth FBGs. Together with the 1310-nm FBGs, this extends the photons' coherence times relative to the pump's to increase the probability that each photon is in a single identical spectral mode \cite{ind_photon}. These narrow-band filters are also crucial for improving tolerance to classical power due to the spectral independence between the SpRS noise photons and the quantum signals. To prevent the detection of C-band photons due to insufficient filter isolation, we cascade two cascaded O-band/C-band WDMs before the FBGs and a 1290-nm DWDM to achieve >190\,dB rejection of C-band light.\\
\indent To perform the BSM, we place PBSs in each output of the 50:50 coupler before the detectors $D_1$ and $D_2$ which are set to project onto orthogonal polarizations $|H\rangle_{D_1}$ and $|V\rangle_{D_2}$. A two-fold coincidence detection then indicates that Alice and Bob's photons were in the $|\Psi^{-}\rangle_{AB}$ Bell state due to HOM interference effects on the other Bell states \cite{PhysRevA.51.R1727}. Due to the non-local correlations of Bob’s entangled photons in the state $|\Psi^{-}\rangle_B$, the destructive measurement of $|\Psi^{-}\rangle_{AB}$ results in the remaining photon at Bob's node being projected onto Alice's initial prepared state $|\psi\rangle_A$ \cite{PhysRevA.51.R1727}. Although teleportation can be performed without projecting onto the $H/V$ basis, Bell state analyzers that measure in the qubit's eigenbasis can double the success probability of a BSM by also measuring the $|\Psi^{+}\rangle$ Bell state if detectors were added to the other ports of our PBSs \cite{Bell_meas}. Including the PBSs thus allows us to investigate BSM nodes most likely to be implemented in future networks. We note that SpRS photons in long-distance fibers are often unpolarized due to polarization mode dispersion \cite{PhysRevApplied.19.044026, Thomas:23} meaning polarizing elements roughly halve the noise rates. This was confirmed in our system by polarization analysis of the noise photons at the BSM node (see supp. materials).\\
\indent Teleportation is evaluated by measuring the polarization of Bob's target photon at detector $D_3$ conditioned on a three-fold coincidence detection between the heralding ($D_0$) and BSM ($D_1$, $D_2$) detectors. All photons are detected by SNSPDs with >90$\%$ efficiency and dark count rates of approximately 100 counts/s. Photon arrival times are determined using a time-to-digital converter and four-fold coincidences are registered when the software delayed time-stamps in each channel fall within a 500-ps time window. The coincidence logic can also be viewed as the classical channel in the teleportation protocol which informs Bob of the BSM result.\\
\indent Since the classical data modulation is uncorrelated with the pulsed quantum sources in the time-domain, the coincidence detection window acts as a temporal filter of the SpRS noise. The temporal and photon-number correlations of our photon-pair sources further help the system’s tolerance to noise by accepting detection events at the receiver only in the time slots when a detection in the heralding arm indicated a photon was transmitted over the noisy channel \cite{herald1}. This can significantly improve the signal-to-noise ratio compared to non-heralded sources that have a high probability of transmitting a vacuum state such as WCSs. In our four-photon teleportation system, both Alice and Bob's pair sources benefit from these correlations.\\ 
\indent The coexistence study is conducted over spooled fiber in a laboratory located at Northwestern University's Evanston campus. The transmission loss over the 30.2-km fiber is 4.9\,dB at the classical signal's 1547.32-nm wavelength and the 1290-nm quantum signals have a total loss of 10.1\,dB (5.1\,dB and 5.0\,dB for Alice and Bob's photons, respectively). At Charlie, the bypassing of the BSM by the classical signal using WDMs adds an additional 1.2\,dB insertion loss. However, the classical transceiver is located at Northwestern's Chicago campus which is connected to Evanston by a 24-km deployed fiber pair and thus operates over a total link distance of 30.2 + 48\,km = 78.2\,km. Due to the high loss over the full link (22.8\,dB) and an initial output power of -9\,dBm, amplification is needed to meet the classical system's $-18$\,dBm minimum received power requirements. We achieve this using erbium doped fiber amplifiers (EDFAs) placed just before multiplexing to co-propagate with Alice's photon. The minimum power launched into the 30.2-km fiber to operate the 400-Gbps channel is $P_{\text{min}} = -3$\,dBm (0.5\,mW) for the receiver located 30.2\,km + 24\,km away from the EDFA, and -10.7\,dBm (85\,$\mu$W) if the classical system operated solely over the 30.2-km link. However, we explore elevated powers using the EDFAs to evaluate whether teleportation can coexist with more hostile classical systems such as ultra high-power long-distance DWDM conventional communications.

\begin{figure*}[!t]
\centering
\includegraphics[width=\linewidth]{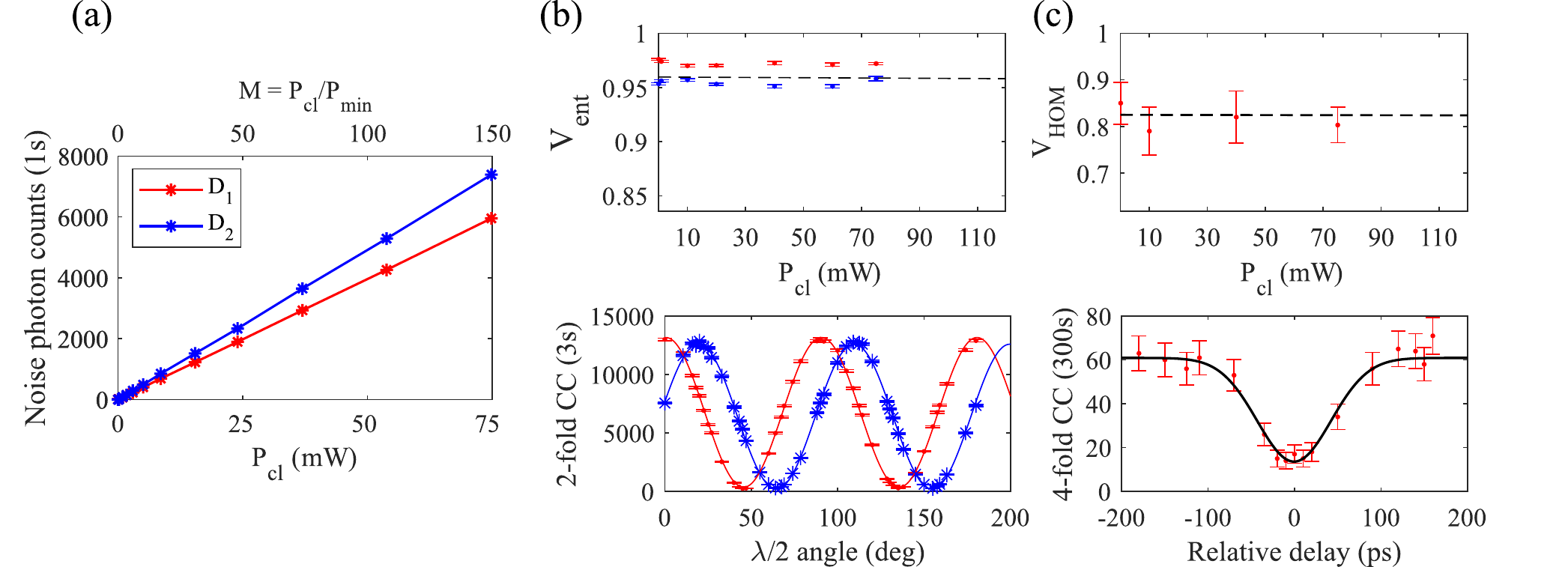}
\caption{(a) SpRS noise photon count rate in the Bell state measurement (BSM) detectors $D_1$ and $D_2$ as a function of the classical launch power $P_{\text{cl}}$ and the ratio $M$ to the minimum power $P_{\text{min}}$ needed to operate a single 400-Gbps channel. (b) Bob's entanglement distribution visibility $V_{\rm ent}$ over 15.0\,km to the BSM node versus $P_{\text{cl}}$ (top) and the corresponding two-fold coincidence count (CC) non-local interference fringe measured in the vertical (red) and anti-diagonal (blue) bases for $P_{\text{cl}}$ = 74\,mW (bottom). (c) Hong-Ou-Mandel interference visibility $V_{\rm HOM}$ over 30.2\,km between heralded photons from Alice and Bob's sources versus $P_{\text{cl}}$ (top) and the corresponding four-fold coincidence interference fringe when $P_{\text{cl}}$ = 74\,mW (bottom). The dashed lines show the predictions of our theoretical model \cite{inprep} (see supp. materials).}
\label{fig:fig2}
\end{figure*}

\section{Experimental results}
\subsection{Noise rates, entanglement distribution, and Hong-Ou-Mandel interference}

We now evaluate the performance of our quantum system. First, we characterize how much noise is introduced into the detectors $D_1$ and $D_2$ as a function of the classical power $P_{\text{cl}}$ that is multiplexed into the 30.2-km fiber. Fig. \ref{fig:fig2}(a) shows the single photon count rates as we increase $P_{\text{cl}}$, where the SNSPD dark count rates have been subtracted. The noise rates scale linearly as 79.0\,counts/s/mW and 97.9\,counts/s/mW in $D_1$ and $D_2$, respectively, the difference is due to slightly unequal efficiencies after the 50:50 coupler. Since a single 400-Gbps channel can operate with as low as $P_{\text{cl}} = 0.5$\,mW, the system can operate with nearly no noticeable noise increase above the SNSPD dark count rate. Fig. \ref{fig:fig2}(a) also shows the ratio $M= P_{\text{cl}}/P_{\text{min}}$, which is often used to roughly estimate the maximum number of channels $M$ that $P_{\text{cl}}$ could support across multiple WDM channels (e.g., \cite{valivarthi_measurement-device-independent_2019, berrevoets_deployed_2022}) assuming each operates at $P_{\text{min}}$ and has similar SpRS properties (see supp. materials).\\
\indent As teleportation requires quantum entanglement as a resource and the ability to have indistinguishable photons for the BSM, we independently characterize the quality of Bob's entanglement distribution to Charlie and HOM interference as we vary $P_{\text{cl}}$. We set Alice and Bob's mean photon pairs per pulse and polarization qubit $\mu$ generated inside each PPLN waveguide to approximately $\mu_A$ = $0.018$ and $\mu_B$ = $0.013$, respectively, which are chosen to balance rates with reduced performance due to multi-pair emission during SPDC \cite{Takeoka_2015}.\\
\indent Fig. \ref{fig:fig2}(b) shows how Bob's entanglement source's non-local two-photon interference fringe visibility in the vertical and anti-diagonal bases is impacted after distributing one photon over 15.0\,km to Charlie as we increase the C-band signal's power launched into the full 30.2-km link. The visibility is defined as $V_{\rm ent}=(R_{\rm max}-R_{\rm min})/(R_{\rm max}+R_{\rm min})$, where $R_{\rm max (\rm min)}$ are the maximum (minimum) two-fold coincidence counts of the interference fringe. Fig. \ref{fig:fig2}(b) shows no noticeable degradation of $V_{\rm ent}$ up to $P_{\text{cl}} =$ 74\,mW, in which we measure $V_{\rm ent,V}= 97.5 \pm 0.1 \%$ and $V_{\rm ent,A} = 95.3 \pm 0.2 \%$. All values are determined by raw data without accidental count subtraction and error bars are calculated via the Monte-Carlo method assuming Poisson photon counting statistics. These results indicate that entanglement distribution is well preserved and that visibility is mainly limited by multi-pair emission during SPDC. These values are also well above the $1/\sqrt{2}$ bound for demonstrating the non-local nature of quantum entanglement \cite{clauser_proposed_1969} and more than sufficient to use for quantum teleportation.\\
\indent Next, we characterize HOM interference between heralded photons from Alice and Bob's sources as we increase $P_{\text{cl}}$. To ensure polarization indistinguishably, we align the PBSs in the BSM setup to instead project onto $|H\rangle_{D_1}|H\rangle_{D_2}$ and prepare Alice and Bob's heralded photons in the state $|H\rangle$. We then record four-fold coincidences as we vary the relative time-of-arrival of each photon using Alice's variable optical delay. The quality of the interference is determined by the visibility of the fringe $V_{\rm HOM}=(R_{\rm max}-R_{\rm min} )/R_{\rm max}$, where $R_{\rm max (min)}$ are the maximum (minimum) four-fold coincidence rates of the HOM interference fringe. The results are shown in Fig. \ref{fig:fig2}(c). We measure a visibility of $V_{\rm HOM} = 82.9 \pm 4.5 \%$ without the classical source, which is comparable to the $V_{\rm HOM} = 83.1 \pm 1.4\%$ visibility we obtain without distribution over the 30.2-km fiber (see supp. materials). We measure $V_{\rm HOM} = 80.3 \pm 3.8 \%$ with $P_{\text{cl}}$ set to 74\,mW, which shows that interference is well preserved in the presence of high-power C-band light but is still limited by the imperfect spectral and single-photon purity of our sources. Each visibility is well above the classical bound of 50\% \cite{RevModPhys.58.209}, indicating our ability to demonstrate non-classical interference across the range of the tested coexisting C-band powers.

\subsection{Quantum teleportation coexisting with high-rate classical communications}

\begin{figure}[!b]
\centering
\includegraphics[width=2.8in]{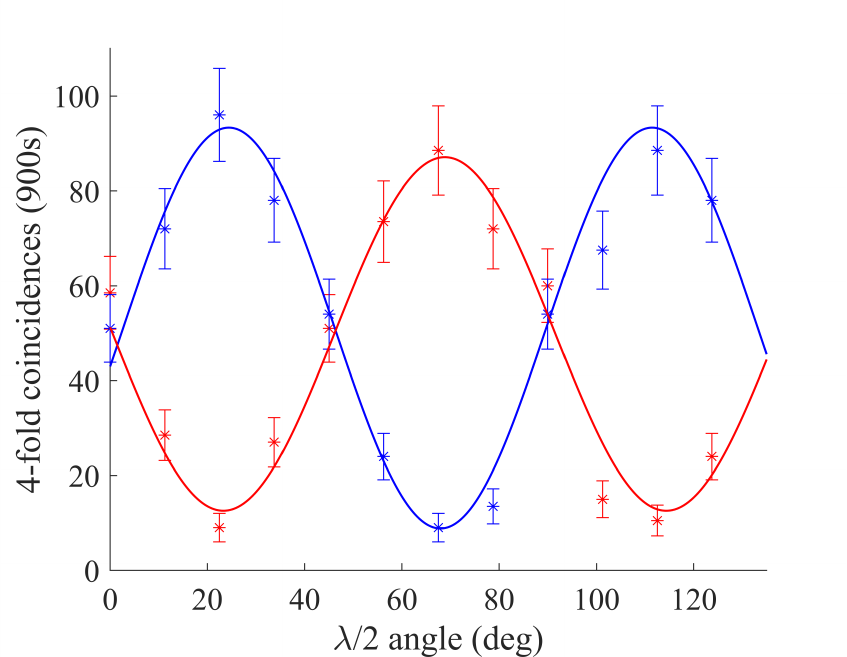}
\caption{Four-fold coincidence counts as we scan Bob's polarization analyzer setting around the Bloch sphere for Alice transmitting $|{D}\rangle$ (blue) and $|A\rangle$ (red) when the launch power of the 400-Gbps C-band classical signal into the 30.2-km fiber is set to 74\,mW.}
\label{fig:fig3}
\end{figure}

\begin{figure*}[!h]
\centering
\includegraphics[width=\linewidth]{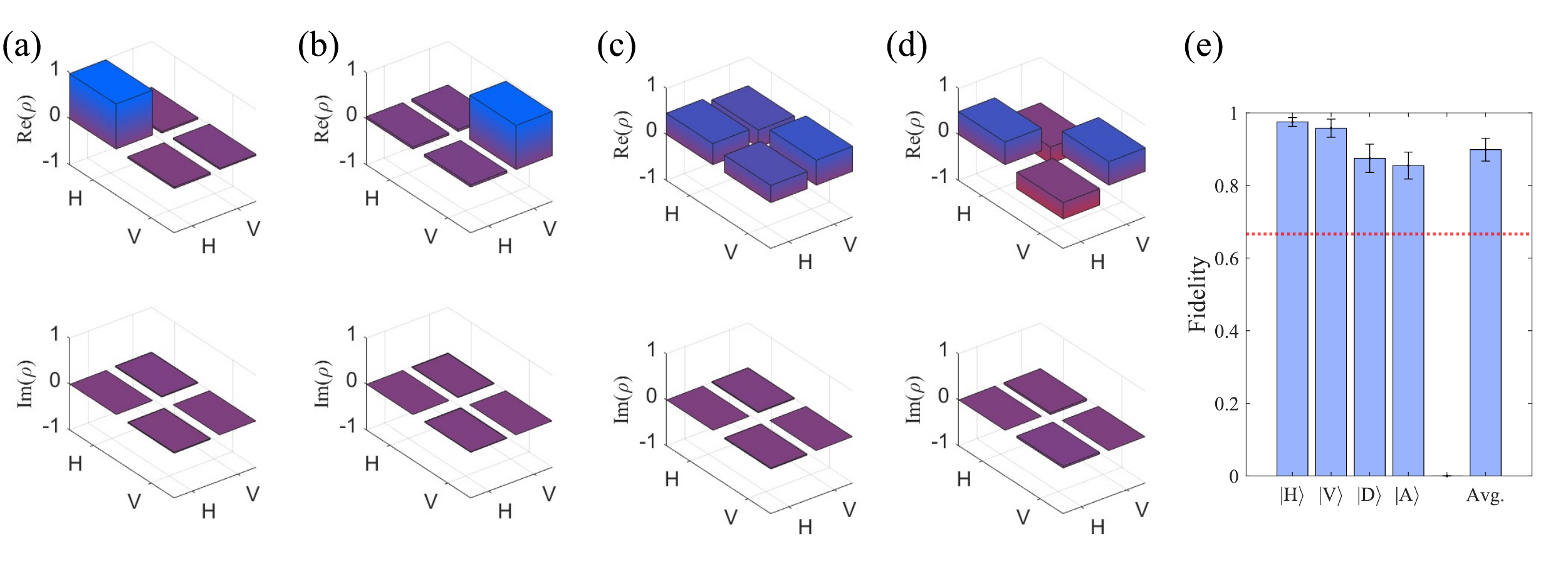}
\caption{Real and imaginary components of the density matrices obtained by single-qubit quantum state tomography on Bob's target photon, conditioned on a successful BSM, whilst the launch power of the 400-Gbps C-band signal into the 30.2-km fiber is set to 74\,mW. (a-d) show the results for Alice transmitting $|\psi\rangle_A = |H\rangle$, $|V\rangle$, $|D\rangle$, and $|A\rangle$, respectively. (e) The fidelity of Bob's density matrix to Alice's ideally prepared qubit and the calculated average fidelity. The dotted line shows the $F = 2/3$ limit for the highest fidelity achievable using only classical physics-based methods \cite{PhysRevLett.74.1259}.}
\label{fig:fig4}
\end{figure*}

Looking at the performance of entanglement distribution and HOM interference, we find that the key operations underlying quantum teleportation are well maintained at high C-band power levels. To demonstrate that this translates to the ability to perform teleportation, we evaluate the teleportation of various qubits while simultaneously transmitting 74\,mW of C-band classical power.\\
\indent Fig. \ref{fig:fig3} shows four-fold coincidence counts as we scan Bob's polarization analyzer setting around the Bloch sphere for Alice transmitting $|{D}\rangle = \frac{1}{\sqrt{2}}(|{H}\rangle +|{V}\rangle)$ and $|{A}\rangle = \frac{1}{\sqrt{2}}(|{H}\rangle -|{V}\rangle)$. We observe for each case a sinusoidal fringe that resembles Alice's prepared qubit with a maximum four-fold rate of $\sim$0.09 counts per second. We obtain fringe visibilities of $V_D = 81.3 \pm 5.4 \%$ and $V_A = 74.7 \pm 4.7 \%$, both values being above the $V= 1/3$ classical limit \cite{PhysRevLett.74.1259}.\\
\indent Next, we perform  maximum likelihood quantum state tomography \cite{ALTEPETER2005105} on Bob's target photon to reconstruct the density matrix $\rho_B$ of his quantum state conditioned on a BSM detection. Fig. \ref{fig:fig4} shows $\rho_B$ for Alice transmitting the states $|H\rangle$, $|V\rangle$, $|D\rangle$, and $|A\rangle$. The fidelity of $\rho_B$ to Alice's ideal qubit $|\psi\rangle_A$ is determined by calculating $F_{\psi} = \langle \psi_A|\rho_B|\psi_A \rangle$. State transfer in the $H/V$ basis does not require HOM interference due to the projection onto the qubit eigen-states in the Bell state analyzer \cite{basso_basset_quantum_2021}, where raw fidelity is mainly limited from unity due to multi-photon pair degradation. In this basis, we measure $F_{H}= 97.5\pm 1.2\%$ and $F_V = 95.8\pm 2.5\%$ for $|{\psi}\rangle_A = |{H}\rangle$ and $|{V}\rangle$, respectively. However, transferring coherent superpositions of $|H\rangle$ and $|V\rangle$ requires interference and thus exhibit more decoherence due to an imperfect spectro-temporal indistinguishability between Alice and Bob's photons. For Alice transmitting $|{\psi}\rangle_A = |{D}\rangle$ and $|{A}\rangle$, we obtain fidelities of $F_D = 87.5\pm 3.9\%$ and $F_A= 85.5 \pm 3.7\%$, respectively. Since the SpRS noise is unpolarized and independent of Alice’s prepared qubit, assuming symmetry of the remaining equatorial states we obtain an average fidelity for an arbitrary qubit on the Bloch sphere of $F_{\rm avg}=\frac{1}{3} F_{\text {poles}}+\frac{2}{3} F_{\text {equator }}= 89.9 \pm 3.1\%$. Furthermore, we measure $F_{\rm avg} = 90.8 \pm 0.8\%$ without sending the photons over the 30.2-km link (see supp. materials), demonstrating that there is minimal degradation of fidelity from distribution over the fiber nor the inclusion of the 400-Gbps signal. The state transfer fidelity of each qubit is shown in Fig. \ref{fig:fig4}(e). All are well above the $F = 2/3$ limit for classical-based methods \cite{PhysRevLett.74.1259}, demonstrating that non-classical teleportation can be achieved alongside high-rate conventional communications.

\section{Discussion and outlook}

In this paper, teleportation was performed alongside $M\approx$150 times higher classical powers than necessary for error-free 400-Gbps communications over the 78-km fiber and $M\approx$870 times greater for a 30.2-km link. Thus, WDM communications could have easily reached many Tbps aggregate data rates without significant impact on teleportation fidelity. The classical power was also on the order of the highest achieved in state-of-the-art studies using direct transmission of quantum states \cite{kumar_coexistence_2015,Mao:18,Geng:21,gavignet_co-propagation_2023,Thomas:23} and indicates the potential for teleportation-based applications operating in the classical backbone fiber infrastructure \cite{Mao:18}.\\ 
\indent Our results also indicate the potential for integrating even more advanced quantum operations. For example, simple modifications to Alice's photon pair source would generate entangled photon pairs and result in entanglement swapping \cite{PhysRevLett.71.4287}, entangling initially independent photons at Alice and Bob's nodes after the BSM. Based on this similarity, it is reasonable to conclude that swapping alongside high-rate classical data is achievable. Further, midpoint BSMs are important for applications such as networking with a central quantum processor \cite{sun_quantum_2016}, quantum repeaters \cite{PhysRevLett.81.5932}, amongst others. Although these will require more capabilities beyond those demonstrated here (e.g., quantum memories, purification, etc.), our results are promising and could be useful for the ongoing development of this technology.\\
\indent Although O-band teleportation networks show promise for high-fidelity integration, C-band quantum networking is usually preferred in quantum-dedicated fiber due to the lowest loss in silica fibers. Unfortunately, C-band channels suffer from orders of magnitude higher SpRS generation rates from C-band classical light \cite{chapuran_optical_2009,Thomas:23}. CV quantum encoding can help address this issue \cite{kumar_coexistence_2015} but is much more fragile to loss compared to discrete-variables which limits its reach. One intriguing possibility is to use high-dimensional entanglement \cite{hyper2024}, but has yet to be fully investigated in high-power regimes. Using our theoretical model \cite{inprep}, we predict C-band teleportation fidelity would be significantly degraded with mW-level coexisting C-band power (see supp. materials) but would improve our fiber transmission loss by $\sim 0.15$ dB/km. This has relevance for research towards long-distance quantum backbone networks since conventional backbone fibers can have $\sim 100$\,mW aggregate power. Further, WDM bandwidth may become limited as some conventional networks push towards fully populating the C- and L-bands \cite{classback}, which also increases the difficulty of avoiding channel cross-talk and optical amplifier noise. One route to alleviate lower O-band quantum rates is to use wavelength multiplexing, which has been shown successful in allowing multi-channel entangled photons populating most of the O-band co-propagating with high-power C- and L-band WDM channels \cite{Thomas:23}. In any case, future research into novel methods to protect quantum fidelity from strong background noise would be invaluable.\\
\indent Some teleportation-based applications require storing photons in quantum memories (QMs), for example to store target photons long enough to receive a signal encoding the BSM result and apply the corresponding unitary transformation to recover Alice's exact state before usage. In our experiment we performed \textit{a posteriori} teleportation \cite{apriori} since Bob's photon is detected before the BSM occurs. Interestingly, quantum mechanics predicts the same correlations irrespective of measurement order. However, attempting to store photons over a long duration can increase photon loss which may complicate the dependence on measurement order in real-world settings having appreciable background noise. If Bob's target channel is completely noiseless, our model predicts that adding pure loss to his target photon's channel does not further degrade the fidelity set by the noise level in the BSM detectors. However, this no longer holds if there is some noise count probability in Bob's detector. In our case we predict our low dark count SNSPDs could tolerate roughly 60\,dB loss in the target channel, which is more than manageable using spooled fiber to delay measurement. However, other approaches such as those using quantum frequency conversion to interface telecom wavelength photons with particular QMs can have notable noise-levels \cite{QFC1} that place more stringent requirements on Bob's channel efficiency (see supp. materials).\\
\indent It is interesting to consider whether teleporting quantum information has any unique benefits in quantum-classical networking compared to direct transmission. It has been shown that using teleportation and BSM operations periodically along a long-distance fiber can increase the possible distances for quantum applications limited by detector dark count noise \cite{relay}, however this may not necessarily apply for SpRS noise generated by classical light and depends on a vast number of parameters such as wavelength channel selections, fiber lengths, co- or counter-propagating signals, and optical amplification for long-distance classical communications. We predict that in some cases there is no advantage if the system has high SpRS noise, but using the methods applied here can help recover some improvements over long-distances and could in principle operate over 100’s of kms, albeit at impractically low quantum rates (see supp. materials). A full analysis of the various scenarios to consider will be pursued in future work. We will also investigate teleportation using a WCS source at Alice's node \cite{valivarthi_quantum_2016,valivarthi_teleportation_2020, QFC2, shen_hertz-rate_2023} as well as the consequences of multiplexing classical communications with entanglement swapping, quantum repeaters, and multi-photon entanglement.\\
\indent In regards to real-world deployment, classical light-based signals are often used in quantum networking for synchronizing distant pulsed quantum sources and detectors to a common clock, communicating measurement results (e.g., BSM results), and monitoring environmental disturbances. Similarly, multiplexing these signals can enhance functionality without sacrificing potentially limited fiber resources \cite{kapoor_picosecond_2023, Burenkov:23, NIST_coex, Qwrap}. Since these can typically operate at sub-mW powers, our results show that including these in our system would have negligible impact if allocated to the C- or L-bands.\\
\indent In conclusion, we have demonstrated quantum state teleportation over a 30.2-km fiber that is populated with high-power 400-Gbps conventional data traffic. By employing various methods to suppress SpRS noise, teleportation fidelity was well maintained alongside elevated classical powers capable of transmitting many Tbps aggregate data rates. We further investigated multiple key questions for optimizing teleportation-based applications and identified challenges facing the deployment of future quantum networks. Altogether, this work demonstrates a significant step towards ensuring complex multi-photon/multi-node quantum network applications can be realized anywhere in the existing fiber infrastructure. 

\begin{backmatter}
\bmsection{Funding} 
This work is funded by Subcontract No. 664603 from Fermi Research Alliance, LLC (FRA) to Northwestern University issued under Prime Contract No. DE-AC02-07CH11359 between FRA and the U.S. Department of Energy (DOE). Although the work is supported by the DOE’s Advanced Scientific Computing Research Transparent Optical Quantum Networks for Distributed Science program, no government endorsement is implied.

\bmsection{Acknowledgments} The authors would like to acknowledge the support of StarLight International/National Communications Exchange Facility and Ciena Corporation. We further thank the members of the Advanced Quantum Networks for Scientific Discovery (AQNET-SD) project, which is a collaboration between Northwestern University, Caltech, Fermilab, Argonne National Laboratory, Jet Propulsion Laboratory, and University of Illinois Urbana-Champaign. We also thank Gina Talcott for help in the editing process. 

\bmsection{Disclosures} The authors declare no conflicts of interest.

\bmsection{Data Availability Statement} Data may be obtained from the authors upon reasonable request. 

\end{backmatter}

% Bibliography
\bibliography{refOPTICAPAPER}
%%%%%%%%%% If preparing manually:
% \begin{thebibliography}{1}
% \newcommand{\enquote}[1]{``#1''}

% \bibitem{Zhang:14}
% Y.~Zhang, S.~Qiao, L.~Sun, Q.~W. Shi, W.~Huang, L.~Li, and Z.~Yang,
%   \enquote{Photoinduced active terahertz metamaterials with nanostructured
%   vanadium dioxide film deposited by sol-gel method,}
%   {\protect\JournalTitle{Optics Express}} \textbf{22}, 11070--11078 (2014).

% \bibitem{Optica}
% {Optica}, \enquote{{Optica Publishing Group},}
%   \url{http://www.opg.optica.org}.

% \bibitem{FORSTER2007}
% P.~Forster, V.~Ramaswamy, P.~Artaxo, T.~Bernsten, R.~Betts, D.~Fahey,
%   J.~Haywood, J.~Lean, D.~Lowe, G.~Myhre, J.~Nganga, R.~Prinn, G.~Raga,
%   M.~Schulz, and R.~V. Dorland, \enquote{Changes in atmospheric consituents and
%   in radiative forcing,} in \enquote{Climate Change 2007: The Physical Science
%   Basis. Contribution of Working Group 1 to the Fourth assesment report of
%   Intergovernmental Panel on Climate Change,}  S.~Solomon, D.~Qin, M.~Manning,
%   Z.~Chen, M.~Marquis, K.~B. Averyt, M.~Tignor, and H.~L. Miler, eds.
%   (Cambridge University Press, 2007).

% \end{thebibliography}

\end{document}